\documentclass[a4paper,fleqn,usenatbib]{mnras}
\usepackage{newtxtext,newtxmath,hyperref}
\usepackage[T1]{fontenc}
\usepackage{ae,aecompl}
\usepackage{graphicx}	% Including figure files
\usepackage{amsmath}	% Advanced maths commands
\usepackage{amssymb}	% Extra maths symbols
\usepackage{url}
\usepackage{bm}

\newcommand{\thee}{\ensuremath{{\theta}_{\rm{e}}}}
\newcommand{\theeg}{\ensuremath{{\theta}_{\rm{e}\gamma}}}

\newcommand{\gr}{$\gamma$-rays }

\newcommand{\ME}{\ensuremath{m_{\rm{e}}}}

\newcommand{\appropto}{\mathrel{\vcenter{
  \offinterlineskip\halign{\hfil$##$\cr
    \propto\cr\noalign{\kern2pt}\sim\cr\noalign{\kern-2pt}}}}}

\title[Constraints on the  emitting region  of the  gamma-rays observed  in GW170817]{Constraints on the  emitting region  of the  gamma-rays observed  in GW170817}

\author[Matsumoto,  Nakar \&  Piran]{Tatsuya Matsumoto,$^{1,2,3}$\thanks{E-mail: tatsuya.matsumoto@mail.huji.ac.il}
Ehud Nakar,$^{4}$
and Tsvi Piran$^{1}$
\\
$^{1}$Racah Institute of Physics, Hebrew University, Jerusalem, 91904, Israel\\
$^{2}$Department of Physics, Graduate School of Science, Kyoto University, Kyoto 606-8502, Japan\\
$^{3}$JSPS Research Fellow\\
$^{4}${The Raymond and Beverly Sackler School of Physics and Astronomy, Tel Aviv University, Tel Aviv 69978, Israel}
}
\pubyear{2018}

\begin{document}
\label{firstpage}
\pagerange{\pageref{firstpage}--\pageref{lastpage}}
\maketitle

\begin{abstract}
The gravitational waves from the neutron star merger event GW170817 were accompanied by an unusually weak short GRB 170817A, by an optical/IR macronova/kilonova and by a long lasting radio to X-rays counterpart. While association of short GRBs with mergers was predicted a long time ago, the  luminosity of this prompt $\gamma$-ray emission was  weaker by a few orders of magnitude than all known  previous sGRBs and it was softer than typical sGRBs. This raise the question whether the $\gamma$-rays that we have seen were a regular sGRB viewed off-axis. 
We revisit this question following recent refined analyses of the $\gamma$-ray signal and the VLBI observations that revealed the angular structure of the relativistic outflow:
observing angle of $\sim 20^\circ$,  a narrow jet with core  $\lesssim 5^\circ$  and $E_{\rm iso} > 10^{52}$ ergs.
We show here that: (i) The region emitting the observed \gr must have been moving with a Lorentz factor  $\Gamma \gtrsim 5$; (ii)  The observed \gr were not
``off-axis" emission (viewing  angle $> 1/\Gamma$)  emerging from the core of the jet, where a regular sGRB was most likely produced; 
(iii)  The $\gamma$-ray emission region was either ``on-axis" (at an angle $< 1/\Gamma$) or if it was ``off-axis" then the observing  angle must have been  small ($< 5^\circ$) and  the on-axis emission from this region was too faint and too hard to resemble a regular sGRB. 
\end{abstract}


\section{INTRODUCTION}\label{intro}

One of the puzzling questions concerning the electromagnetic (EM) counterparts of the gravitational waves (GWs) from the binary neutron star merger GW170817 \citep{Abbott+2017c} is the origin of the \gr observed 1.7 s following the merger \citep{Goldstein+2017,Savchenko+2017}.  Binary neutron star mergers were suggested already 30 years ago as the progenitors of short gamma-ray bursts (sGRBs) \citep{Eichler+1989}. Many indirect evidence supporting this suggestion have been found since then (see \citealt{Nakar2007} and \citealt{Berger2014} and references therein). Therefore the first reaction to the observed $\gamma$-rays was that we observed a regular sGRB \citep{Goldstein+2017}.  

However, the $\gamma$-ray signal was unlike any sGRB seen before. Most significant was its luminosity. It is fainter by about three orders of magnitude than the faintest sGRB observed so far. In addition both the spectral hardness and the spectral evolution were uncommon (soft peak energy and two distinct spectral components, the second one consistent with being a blackbody). One of advocated solutions was that GW170817 produced an ultra-relativistic jet, which emitted a regular sGRB, as seen by an observer that is within the jet cone, and that we are out of the jet cone seeing the {\it same $\gamma$-rays} off-axis \citep{Goldstein+2017,Murguia-Berthier+2017b,Ioka&Nakamura2018}.
Namely, the only difference between the \gr that we observed and the emission (presumably a regular sGRB) seen by an ``on-axis" observer was the different Lorentz boosts from the source to the observers.

A closer inspection of the $\gamma$-ray signal using compactness arguments 
and the late (16 and 9 days) onset of the radio and X-ray afterglows 
\citep{Alexander+2017,Hallinan+2017,Kim+2017,Mooley+2018,Resmi+2018,Nynka+2018,DAvanzo+2018,Alexander+2018,Dobie+2018,Corsi+2018,Mooley+2018b,Haggard+2017,Margutti+2017,Troja+2017,Troja+2018,Margutti+2018,Ruan+2018}
suggested that this is not the case.
Namely, even if  GW170817  was associated with  a regular sGRB along its rotation axis we did not see these $\gamma$-rays and the origin of the $\gamma$-rays that we observed is different \citep{Kasliwal+2017,Gottlieb+2018,Gottlieb+2018b,Troja+2017,Granot+2017}.

The main suggested explanation for the \gr we have seen was that 
the outflow contained a component with a lower Lorentz factor and a lower isotropic equivalent energy than that of a typical sGRB jet.  This component produced the observed \gr via a different emission mechanism than the one operating in regular GRBs \citep{Kasliwal+2017,Lazzati+2017c,Kathirgamaraju+2018,Gottlieb+2018b,Bromberg+2018,Pozanenko+2018}. 
A natural origin for such an outflow is the cocoon created by the interaction of a relativistic jet with the sub-relativistic merger ejecta, and is expected to produce a relatively wide-angle ($\sim 0.5$ rad) mildly relativistic component \citep{Meszaros&Rees2001,Ramirez-Ruiz+2002,Nakar&Piran2017,Lazzati+2017b,Gottlieb+2018}. Since a mildly relativistic cocoon can arise from  either an emerging or a choked jet \citep{Gottlieb+2018b,Nakar+2018}, the \gr  could not be used to confidently determine the fate of the jet in GW170817.   

Recent  VLBI images  show a super-luminal motion  \citep{Mooley+2018b}. When combined with the radio light curve these
observations suggest that GW170817 involved a narrow and energetic jet along its rotation axis. This jet has emerged from the merger's ejecta (hereafter we denote such a jet as a ``successful" jet) and most likely produced a powerful sGRB that could have been seen by observers along its narrow opening angle. These observations also enabled us to estimate the system's geometry,  our viewing angle, the parameters of the surrounding matter, and the microphysical parameters of the shocks involved.  In addition, \cite{Veres+2018} preformed a refined analysis of the Fermi-GBM data obtaining  time-resolved spectrum for the initial luminous and hard $\gamma$-ray pulse. 

The new  results offer an opportunity to improve the constraints  on the source of the $\gamma$-ray signal and to compare its properties and angular position with those of the jet's core. This is the goal of this paper.
Before beginning we clarify an important terminology that we use here and elsewhere. We consider an outflow moving with a Lorentz factor $\Gamma$. Generally $\Gamma$ is a function of $\theta$ the angle from the axis. We define an on-axis (off-axis) emission as emission coming to the observer from an angle $\Delta \theta < 1/\Gamma$ ($\Delta \theta >1/\Gamma$), {where $\Delta\theta$ is the angular distance between the emitting region and the observer.}

Following a short summary of the observations in \S \ref{sec:obs}, we explore in \S \ref{sec:prompt} the compactness limits on the emitting region. We obtain a lower limit on the Lorentz factor of the emitting region that we see off-axis, and an upper limit on its angle with respect to the line-of-sight as a function of the  on-axis $\gamma$-ray isotropic equivalent luminosity. 
In \S \ref{sec:afterglow}, we turn to the observed radio afterglow and determine the maximal isotropic equivalent kinetic energy of the outflow at any given angle with respect to the line-of-sight, so that it will not overproduce the observed radio signal.
A comparison of this upper limit with the $\gamma$-ray energy needed to produce the observed prompt emission as an off-axis emission outlines the allowed region for the condition within the prompt emitting region. 
In \S \ref{sec:GRBs}, we discuss  the possibility that the emission that we have seen was observed  as a regular sGRB by an on-axis observer, finding that it is highly unlikely.
These findings do not mean that the event was not accompanied by a sGRB. However, we did not observe directly any emission from the region that has produced this sGRB.  The  prompt \gr 
that we observed were most likely produced by a different mechanism \citep[see e.g.][]{Kasliwal+2017,Gottlieb+2018,Gottlieb+2018b}. We discuss 
and summarize the implications of these results in \S \ref{sec:conclusions}.

\section{The Observations}\label{sec:obs}

The $\gamma$-ray detectors on \textit{Fermi} and \textit{INTEGRAL} were triggered $\simeq1.7\,\rm{s}$ after the detection of the GWs \citep{Abbott+2017e,Goldstein+2017,Savchenko+2017}.
The observed \gr showed a first smooth pulse followed by a second softer one.
The spectrum of the first pulse is fitted by the Comptonized model (a power-law with exponential cutoff) with a power-law index of ${\alpha_p}\simeq-0.62\pm0.40$ and a peak energy of $E_{\rm{p}}\simeq185\pm62\,\rm{keV}$.
The latter pulse is fitted by a black body spectrum with a temperature of $k_{\rm{B}}T\simeq10.3\pm1.5\,\rm{keV}$.
Given the distance, the  isotropic energy and luminosity are  estimated as $E_{\rm{\gamma,iso}}\simeq(5.4\pm1.3)\times10^{46}\,\rm{erg}$ and $L_{\rm{\gamma,iso}}\simeq(1.6\pm0.6)\times10^{47}\,\rm{erg\,s^{-1}}$, respectively.

\cite{Veres+2018} carried out a time-resolved spectral analysis on the first pulse.
They find that the peak energy and the luminosity during the first time step are $E_{\rm{p}}\simeq520_{-290}^{+310}\,\rm{keV}$ and $L_{\rm{\gamma,iso}}\simeq2.0_{-0.6}^{+0.6}\times10^{47}{\,\rm{erg\,s^{-1}}}$ at a time-bin of $\delta{t}=0.064\,\rm{s}$. Both decline later. 
As we show below these observed values impose stronger constraints on the conditions within the emitting region than those obtained earlier based on the average flux and energy \citep{Kasliwal+2017,Granot+2017,Ioka&Nakamura2018}. 

VLBI observations on 75 and 230 days revealed a rather compact nebula whose centroid moves across the plane of the sky at an apparent velocity of $\beta_{\rm app} = 4.1 \pm 0.5$, measured in units of the speed of light \citep{Mooley+2018b}. When combined with the rapid decay of the radio flux following the peak at $\sim 150$ d, this result is best explained by a narrow source that moves between days 75 and 230 at $\Gamma \approx \beta_{\rm app}$ at an angle $\Delta \theta \approx 1/\beta_{\rm app}$ with respect to our line-of-sight. Detailed modeling shows  an energetic
 ($E_{\rm k,iso} \gtrsim 10^{52}$ ergs) narrow ($\theta_{\rm j} \lesssim 0.1{\rm\,rad}\simeq5^\circ$) jet at the core. The jet must be surrounded by a wider component with a lower energy,  that is fully consistent with the cocoon driven by the jet. Our angle with respect to the edge of jet core is $\theta_{\rm obs}-\theta_{\rm j} \simeq 0.2-0.4\,\rm{rad} \simeq 11^\circ-22^\circ $ and with respect to the jet symmetry axis  $\theta_{\rm obs}  \simeq 0.25-0.5\,\rm{rad} \simeq 14^\circ-28^\circ$.
Fitting the afterglow data, \cite{Mooley+2018b} find that typical microphysical parameters for a jet with isotropic equivalent energy $E_{\rm k,iso} \sim 10^{52}\,\rm ergs$ are $\epsilon_{\rm B} \simeq 10^{-3} $ and $n \simeq 3 \times 10^{-4} \, {\rm cm}^{-3} $ (assuming $\epsilon_{\rm e}\simeq 0.1$). A larger jet energy requires a larger  external density ($n \propto E_{\rm k,iso}$) and a lower magnetic field ($\epsilon_{\rm B} \appropto E_{\rm k,iso}^{-2}$).

\section{Prompt emission constraints on the gamma-ray source}\label{sec:prompt}

As in regular GRBs, compactness   constrains  the Lorentz factor of the emitting region  and its viewing angle \citep{Kasliwal+2017}.
The optical depth in the rest frame is given by \citep{Nakar2007}
\begin{eqnarray}
\tau_{\gamma}\simeq\frac{\sigma_{\rm{T}}N_{\rm{ph}}f}{4\pi{R^2}},
   \label{tau}
\end{eqnarray}
where $\sigma_{\rm{T}}$, $N_{\rm{ph}}$, $f$, and $R$ are the Thomson cross section, total photon number, the fraction of photons which can create pairs, and the size of the emission region, respectively.
There are two limits on the minimum Lorentz factor \citep{Lithwick&Sari2001}.
First, a photon with a typical observed energy $E_{\rm{p}}$ (or the maximum observed energy) can annihilate a photon whose energy is larger than $E\gtrsim(\Gamma\ME{c^2})^2/E_{\rm{p}}$, where $\Gamma$, $\ME$, and $c$ are the Lorentz factor of the outflow, the electron mass, and the speed of light, respectively.
For this photon to escape, the optical depth for  pair creation should be smaller than unity (limit A).
A second requirement is that the photons should not be scattered off by pairs that have been  created by photons with $E\gtrsim\Gamma\ME{c^2}$ (limit B).

For GRB 170817A,  limit B imposes the more constraining condition.
For the Comptonized photon spectrum of $dN/dE\propto{E^{\alpha_p}}e^{-({\alpha_p}+2)E/E_{\rm{p}}}$ with $-1<{\alpha_p}<0$, the fraction, $f$,  is given by\footnote{This fraction was approximated by $f\simeq\Gamma^{\alpha_p}{\exp\biggl[-\frac{\Gamma\ME{c^2}}{E_{\rm{p}}/({\alpha_p}+2)}\biggl]}$ in \cite{Kasliwal+2017}, which is less accurate than the approximation we use here.}
\begin{eqnarray}
f&\simeq&\frac{\int_{\Gamma\ME{c^2}}^\infty{dE\frac{dN}{dE}}}{\int_{0}^\infty{dE\frac{dN}{dE}}}  \simeq {\bf \Gamma}\biggl(\alpha_{p}+1,\frac{\Gamma\ME{c^2}}{{E_{\rm{p}}/({\alpha_p}+2)}}\biggl)\nonumber\\
&\simeq&\biggl(\frac{\Gamma\ME{c^2}}{E_{\rm{p}}/({\alpha_p}+2)}\biggl)^{{\alpha_p}}{\exp\biggl[-\frac{\Gamma\ME{c^2}}{E_{\rm{p}}/({\alpha_p}+2)}\biggl]},
\end{eqnarray}
where 
we consider a normalized distribution so that ${\int_{0}^\infty{dE\frac{dN}{dE}}} \equiv 1$ 
and we approximate  the incomplete gamma function, ${\bf \Gamma}({\alpha_p}+1,x)\simeq{x^{{\alpha_p}}}e^{-x}$ for $x\gg1$.
The photon number and the emission size are given by $N_{\rm{ph}}\simeq{L_{\rm{\gamma,iso}}\delta{t}}/E_{\rm{p}}$ and $R=c\Gamma^2\delta{t}$, where $L_{\rm{\gamma,iso}}$ and $\delta{t}$ are the isotropic gamma-ray luminosity and the duration, which is equal to the pulse duration for GRB 170817A.
Eq. \eqref{tau} is rewritten as
\begin{eqnarray}
&&\tau_{\gamma}\simeq7.2\times10^{11}\,\frac{L_{\rm{\gamma,iso,51}}}{({\alpha_p}+2)\delta{t}_{-1}}\times 
\nonumber\\
&&\biggl(\frac{\ME{c^2}}{E_{\rm{p}}/({\alpha_p}+2)}\biggl)^{{\alpha_p}+1}\Gamma^{{\alpha_p}-4}\exp\biggl[-\frac{\Gamma\ME{c^2}}{E_{\rm{p}}/({\alpha_p}+2)}\biggl].
   \label{tau2}
\end{eqnarray}
Hereafter, we adopt the convention $Q_x=Q/10^x$ (cgs units).
This equation is derived for an on-axis observer.

We consider now an observer located at $\theta_{\rm{obs}}$ (all angles are measured relative to the rotation axis of the system) and an 
off-axis $\gamma$-ray emitting region  that is located at an angle $\theta_{\rm{e}\gamma}$ such that 
$\theta_{\rm{e}\gamma}<\theta_{\rm{obs}}$ and $q\equiv\Gamma(\theta_{\rm{obs}}-\theta_{\rm{e}\gamma}) \gg 1$.
We denote  observables for this  off-axis observer with a prime.
The photon energy is given by 
\begin{eqnarray}
\frac{E_{\rm{p}}}{E_{\rm{p}}^\prime}=\frac{1-\beta\cos(\theta_{\rm{obs}}-\theta_{\rm{e}\gamma})}{1-\beta}\simeq{q^2} \ ,
   \label{e_peak}
\end{eqnarray}
where $\beta$ is the outflow velocity measured in units of the speed of light.

The isotropic energy is transformed in a more complicated way depending on the viewing angle \citep{Kasliwal+2017,Granot+2017,Ioka&Nakamura2018}:
\begin{eqnarray}
\frac{E_{\gamma,\rm{iso}}}{E_{\rm{\gamma,iso}}^\prime}\equiv{\cal{A}}\simeq\begin{cases}
q^4&\text{;\,$\theta_{\rm{obs}}-\theta_{\rm{e\gamma}}\ll\theeg$\,\ \ \ \ \ \ \ \  \ (\textit{i})},\\
q^6(\Gamma{\theta_{\rm e\gamma}})^{-2}&\text{;\,$\theta_{\rm{obs}}-\theta_{\rm{e}\gamma}\gg\theeg>1/\Gamma$\,(\textit{ii})},\\
q^6&\text{;\,$\theeg<1/\Gamma$\,\ \ \ \ \ \ \ \ \ \ \ \ \ \ \ \ \ (\textit{iii})}.
\end{cases}
   \label{e_gamma_iso}
\end{eqnarray}
Cases ($i$) [($ii$)] is when the observer is near [far] the edge of the emitting region (but off-axis). Both hold until the region expands laterally $\Gamma\theta_{\rm{e}\gamma}>1$.
Case ($iii$) is after the lateral expansion. While it is unlikely in the prompt phase it is included for  completeness.

The transformation of durations is less trivial and we need to distinguish between the duration of each emission episode, $\delta t$, and that of the entire observed burst, $\Delta t$. The transformation of the duration of the emission from a single episode is simply 
\begin{eqnarray}
\frac{\delta{t}}{\delta{t}^\prime}\simeq{q^{-2}}.
\label{eq:delta_t}
\end{eqnarray}
In our frame we see only a single pulse. % namely $\Delta t^\prime=\delta{t}^\prime$. 
This pulse may correspond to a single emission episode, in which case an on-axis observer will also see a single pulse with $\Delta t=\delta{t}=\delta{t}^\prime q^{-2}$. However there is another possibility. If each emission episode corresponds to a different part of the outflow (as for example in internal shocks) 
then the time separating two pulses is the same for on-axis and off-axis observers. Since the time between pulses remains invariant in this scenario, while the duration of each pulse is longer for an off-axis observer by a factor of $q^2$, a burst seen off-axis may be much less variable than when it is seen on-axis. Thus, a single pulse seen by an off-axis observer, may be seen as many (up to $q^2$) different pulses by an on-axis observer. These different possibilities are important when we calculate the transformation of the luminosity in Eq. \eqref{tau2}, which corresponds to the luminosity of a single pulse in the frame of an on-axis observer. In one extreme we can assume that there is only a single emission episode and then an on-axis observer sees only a single pulse with a duration $\delta t= \delta t^\prime {q^{-2}}$ that contains all the observed $\gamma$-ray energy. Thus, Eqs. \eqref{e_gamma_iso} and \eqref{eq:delta_t} imply $L_{\rm{\gamma,iso}}=L_{\rm{\gamma,iso}}^\prime {\cal{A}}q^2$. The other extreme option is that there are $\sim q^2$ identical emission episodes, each with a duration that is similar the separation between two pulses as seen by an on-axis observer. In this case the duration of each pulse is  $\delta t= \delta t^\prime {q^{-2}}$, but it contains only a fraction of $q^{-2}$ of the total $\gamma$-ray energy, implying $L_{\rm{\gamma,iso}}=L_{\rm{\gamma,iso}}^\prime {\cal{A}}$ . To conclude,
\begin{eqnarray}
{{L_{\rm{\gamma,iso}}}\over {L_{\rm{\gamma,iso}}^\prime}} = {\cal{A}}{q^p} \ ,
\end{eqnarray}
where $0<p<2$, depending on the (unknown) light curve shape (i.e., variability) seen by an on-axis observer.

Now we can replace the on-axis observables with the off-axis ones in Eq. \eqref{tau2}:
\begin{eqnarray}
\tau_{\gamma}&\simeq&7.2\times10^{7}\,\frac{L_{\rm{\gamma,iso,47}}^\prime}{({\alpha_p}+2)\delta{t}_{-1}^{\prime}}\biggl(\frac{\ME{c^2}}{E_{\rm{p}}^\prime/({\alpha_p}+2)}\biggl)^{{\alpha_p}+1}\nonumber\\
&&\frac{{\cal{A}}q^{p-2{\alpha_p}}}{\Gamma^{4-{\alpha_p}}}\exp\biggl[-\frac{\Gamma\ME{c^2}}{q^2E_{\rm{p}}^\prime/({\alpha_p}+2)}\biggl].
   \label{tau3}
\end{eqnarray}
Note that $q$ here is a function of $\cal{A}$ (or vice versa).

\begin{figure}
\begin{center}
\includegraphics[width=60mm, angle=90]{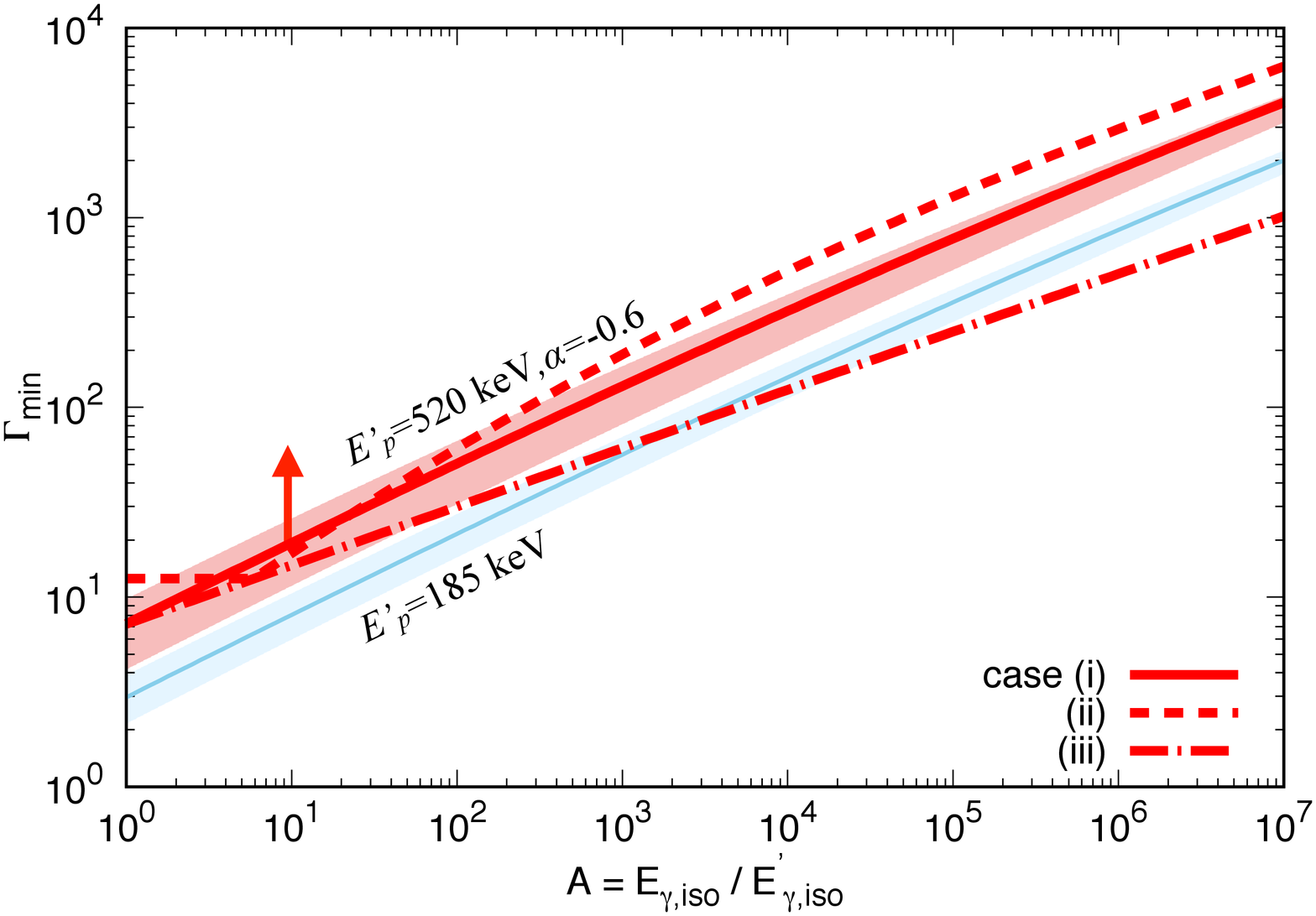}
\caption{Lower limits on the Lorentz factor derived from the compactness argument.
The red solid, dashed, dash-dotted curves correspond to limits imposed by the observed parameters of the first time-bin of \citep{Veres+2018} for cases ($i$)-($iii$), respectively.
For case ($ii$), we use $\theta_{\rm{e}\gamma}=0.08\,\rm{rad}$ \citep{Mooley+2018b}.
The thin blue curve denotes the limit imposed by using the average parameters \citep{Goldstein+2017} for case ($i$) only.
The red and blue shaded regions show the uncertainties resulting from the $\gamma$-ray spectrum.
For case ($ii$), the Lorentz factor has another lower limit, $\Gamma>\theta_{\rm{e\gamma}}^{-1}$ (Eq. \ref{e_gamma_iso}). One can see that for an on-axis observer (${\cal A}=1$) the minimal Lorentz factor is $\sim 5$. }
\label{fig gamma}
\end{center}
\end{figure}

The condition $\tau_{\gamma} \lesssim 1$ yields a lower limit on the Lorentz factor, $\Gamma>\Gamma_{\rm min}(\cal{A})$.
Fig. \ref{fig gamma} depicts $\Gamma_{\rm min}$ for cases ($i$)-($iii$), where for case ($ii$) we have to assume the angular position of the emitting region,  $\theeg$. Following \cite{Mooley+2018b}, we set $\theeg=\theta_{\rm{j}}=0.08\,\rm{rad}$, as an approximate size of the jet core in GW170817. We take the most conservative assumption and set $p=0$, which gives weaker lower limits on $\Gamma$.  
We plot $\Gamma_{\rm min}$ for the parameters given by \cite{Veres+2018} at the luminosity peak, which are the most constraining ($E_{\rm{p}}^\prime=520\,\rm{keV}$, $\alpha_p=-0.6$, $L_{\rm{\gamma,iso,47}}^\prime=2.0$, and $\delta{t}_{\rm{-1}}^\prime=0.64$).
The effect of the uncertainty in the peak energy and luminosity is shown as a shaded region, which is calculated for case (i) using upper and lower boundaries that are given by $(E_{\rm{p}}^\prime,\,L_{\rm{\gamma,iso,47}}^\prime)=(830{\rm{\,keV}},\,2.6)$ and $(230{\rm{\,keV}},\,1.4)$, respectively.
For a comparison we draw a blue curve and the corresponding shaded region that show the limits  for  case ($i$) as derived by using the average observables ($E_{\rm{p}}^\prime=185\,\rm{keV}$, $\alpha=-0.6$, $L_{\rm{\gamma,iso,47}}^\prime=1.6$, and $\delta{t}_{\rm{-1}}^\prime=10$). These values  give, of course, a less stringent  limit.

\begin{figure}
\begin{center}
\includegraphics[width=60mm, angle=90]{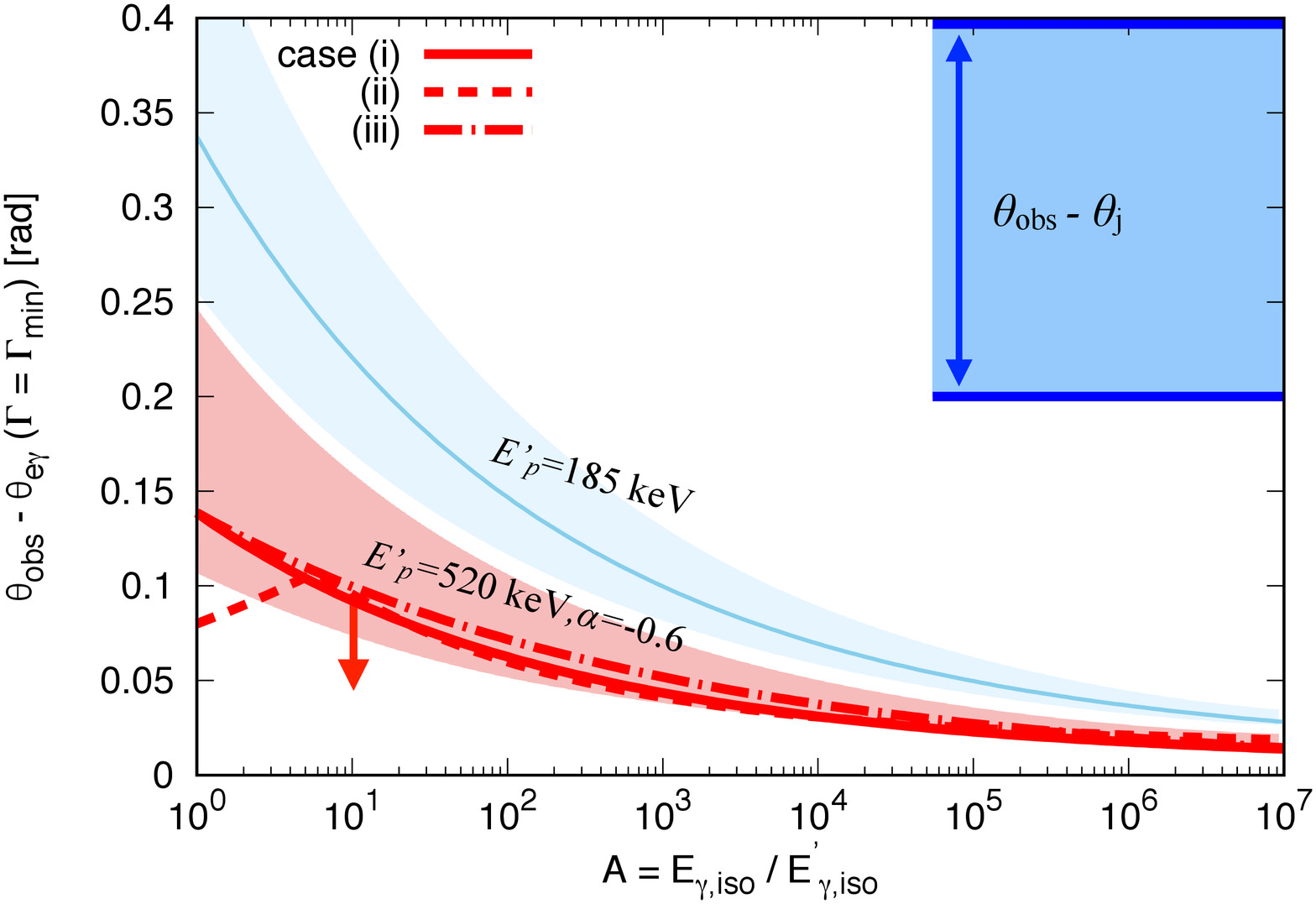}
\caption{Maximal angular distance between the viewing angle $\theta_{\rm{obs}}$ and emitting region  $\theeg$ imposed by the compactness argument.
The  curves are depicted for the same parameters as in Fig. \ref{fig gamma}.
The blue horizontal lines show the angular distance, given by the VLBI analysis \citep{Mooley+2018b},  between the observer and the core of the jet and the value of ${\cal A}$ assuming that the core produced a regular sGRB.}
\label{fig theta}
\end{center}
\end{figure}

The lower limit on the Lorentz factor yields an upper limit on the angular distance between the observer and the emitting region,  $\theta_{\rm{obs}}-\theeg<q/\Gamma_{\rm min}$ (see Eq. \ref{e_gamma_iso}).
Fig. \ref{fig theta} depicts these limits.
In all cases, the angular distance is constrained by $\theta_{\rm{obs}}-\theeg\lesssim0.1\,\rm{rad}$ for ${\cal A} \gtrsim 10$, while  
\cite{Mooley+2018b} find $0.25{\,\rm rad}<\theta_{\rm{obs}}<0.5$ rad (similar lower limits on $\theta_{\rm{obs}}$ were obtained earlier by examination of the GW signal by \citealt{Mandel2018,Finstad+2018}). Case ($ii$), which requires $\theta_{\rm{obs}}-\theeg \gg \theeg$, is inconsistent with this result, implying that if there are significant off-axis effects (i.e., ${\cal A} \gtrsim 10$) then we must be in case ($i$), namely the emitting region is closer to the observer than it is to the jet axis.

Fig. \ref{fig theta} also shows the properties of the sGRB that was presumably emitted by the core of the jet, based on the constraints set by the VLBI observations ($0.2{\,\rm rad}<\theta_{\rm{obs}}-\theeg<0.4$ rad and $E_{\gamma, \rm iso} \sim E_{\rm k,iso} \gtrsim 10^{52}$ ergs, which corresponds to ${\cal A} \gtrsim 10^5$). It is clearly evident that if GW170817 produced sGRB in the direction of the jet core, the \gr that we observed are not this sGRB seen off-axis. In fact, Fig. \ref{fig theta} demonstrates that regardless of the value of ${\cal A}$ the origin of the \gr we observed must be very far from the jet core.

Finally, ${\cal A}=1$ (and $q=1$) corresponds to an on-axis emission (namely $\theta_{\rm{obs}}-\theeg<1/\Gamma$). Therefore plugging ${\cal A}=1$  into Eq. \eqref{tau3} provides an absolute lower limit on the Lorentz factor of the $\gamma$-ray emitting region. Fig. \ref{fig gamma} shows that when considering the uncertainty in $E_{\rm p}$ (the lower edge of the red shaded region in Fig. \ref{fig gamma}) that if the $\gamma$-ray emission was seen on-axis then the emitting region Lorentz factor must  be $\Gamma \gtrsim 5$.  

\section{Afterglow Constraints on the Kinetic energy}\label{sec:afterglow}

The observed afterglow puts upper limits on the isotropic equivalent kinetic energy carried by relativistic material at regions with different angular distances from the observer, i.e. $E_{\rm k,iso}(\theta)$.
The contribution to the afterglow flux from a given region is brighter when its $E_{\rm k,iso}$ is larger. Similarly, for a given value of $E_{\rm k,iso}$, 
the emission from a region at a smaller angular distance is brighter and it peaks earlier.  Therefore, $E_{\rm k,iso}(\theta)$ is limited by the requirement that it does not overproduce the  observed afterglow flux. Since the outflow is expected to have both angular and radial structures, for every region this constraint accounts only for material with initial Lorentz factor that is large enough to contribute to the forward shock at the time of the observations. We find in \S \ref{sec:prompt} that the Lorentz factor of the $\gamma$-ray emitting region is $\gtrsim 5$. At the same time \cite{Mooley+2018b} find that the Lorentz factor of the shock driven by the core of the jet into the circum-merger medium is $\approx 4$ at the time that it starts dominating the emission. Therefore,  in the following analysis  we constrain the energy carried by material with initial Lorentz factor $\gtrsim 5$. This limit is general and it is valid for any given region, regardless of the question whether this region is the source of the \gr or not, and therefore  it is important by itself. However, when combined with the limits imposed by compactness it puts tight constraints on the  properties of the $\gamma$-ray emitting region.

In order to derive the limit on $E_{\rm k,iso}(\Delta\theta)$,  where\footnote{Note that $\thee$ here is the angle of the region that contributes to the afterglow, not to be confused with $\theeg$ defined in \S \ref{sec:prompt}, which is the region that produces the prompt \gr.} $\Delta\theta\equiv\theta_{\rm{obs}}-\thee$, we estimate the  maximal flux that this region generates and the time that this flux is observed, as a function of $E_{\rm k,iso}$. This can be done relatively well (to within an order of magnitude) since the afterglow observations \citep{Mooley+2018b}  provide estimates of   the microphysical parameters $n$ and $\epsilon_{\rm{B}}$  (see \S \ref{sec:obs}). We separate the constraints for material that is moving directly towards us, namely an on-axis material as defined based on its initial Lorentz factor, and material that is initially off-axis, namely at $\Delta\theta >\Gamma_0^{-1}$, where $\Gamma_0$ is the initial Lorentz factor of the material at $\Delta\theta$. The former, on-axis limit, is a function of $\Gamma_0$  and it constrains only the energy carried by material faster than $\Gamma_0$. The constraints on the initially off-axis material is a function of $\Delta\theta$ and it constrains material that moves with an initial $\Gamma_0 >\Delta\theta^{-1}$. Note that we use the term ``initially" since by the time that we see the contribution from each region its Lorentz factor decelerated to the point that $\Gamma \lesssim\Delta\theta^{-1}$ and is therefore on-axis at that time \citep{Nakar&Piran2018}.

We  consider three different possible regimes: (i) initially on-axis ($\Delta\theta <\Gamma_0^{-1}$) (ii) initially near off-axis ($\Gamma_0^{-1}<\Delta\theta \ll \theta_{\rm{e}}$) and (iii) initially far off-axis ($\Gamma_0^{-1}<\theta_{\rm{e}} \ll \Delta\theta$). We compare the emission from  each regime to observations. At $10<t<150$ d, we use 3 GHz observations \citep{Mooley+2018}: $F_\nu(3{\,\rm GHz}, ~t>10 {\rm ~d}) \simeq 13\,{\rm{\mu{Jy}}}(t/10\,{\rm{days}})^{0.78}$. At $2<t<10$ d, we use X-ray \citep{Troja+2017} and 6 GHz upper limits \citep{Hallinan+2017}, assuming that (as predicted by theory) the spectrum is constant during that time, $F_\nu(3{\,\rm GHz},~2<t<10 {\rm ~d}) < 13\,{\rm{\mu{Jy}}}$. Before day 2 there are no effective constraints on the afterglow. The estimates for all cases are based on analytic formulae derived in previous studies, where the normalization is adopted from a semi-analytic code described in \cite{Soderberg+2006} that takes into account the geometrical factors. {Following \cite{Mooley+2018b}, we use the  canonical microphysical parameters:  $n=10^{-4} {\rm~cm^{-3}}$, $\epsilon_e=0.1$, $\epsilon_B=10^{-3}$ and  $p=2.16$.}

\subsection{On-axis ($\Delta\theta = 0$)}
We approximate the on-axis emission as a top-hat jet with an opening angle $\theta_{\rm j}=\theta_{\rm obs}$ and an initial Lorentz factor $\Gamma_0$. The emission from such an outflow peaks at \citep{Sari+1998}:
\begin{eqnarray}
t_{\rm{peak}}\simeq 30 \,{\rm{d}}\,E_{\rm{k,iso,51}}^{1/3}n_{-4}^{-1/3}\left(\frac{\Gamma_0}{5}\right)^{-8/3}\ ,
   \label{peak time1}
\end{eqnarray}
and its 3GHz flux at that time is  
 \begin{eqnarray}
F_{\nu,\rm{peak}}(\rm{3~GHz})&\simeq& 400{\,\mu \rm{Jy}\,}E_{\rm{k,iso,51}}n_{-4}^{\frac{p+1}{4}}\epsilon_{\rm{B,-3}}^{\frac{p+1}{4}}\nonumber\\
&&\epsilon_{\rm{e},-1}^{p-1}\left(\frac{\Gamma_0}{5}\right)^{2(p-1)} d_{\rm{40Mpc}}^{-2} \ .
   \label{maximum flux1}
\end{eqnarray}
A comparison to the observations shows that the isotropic equivalent energy carried by material that moves at $\Gamma_0 >5$ towards  us is $E_{\rm{k,iso}}(\Gamma_0 >5) \lesssim 3 \times 10^{49}$ erg. This isotropic equivalent energy is about three orders of magnitude smaller than that of the jet core, but it is still more than enough to allow material that moves towards the observer to emit the observed \gr. Applying this constraint to material with lower values of $\Gamma_0$, we find for example $E_{\rm{k,iso}}(\Gamma_0 \approx 3) \lesssim 6 \times 10^{50}$ erg. 

\subsection{Near off-axis ($\Gamma_0^{-1}<\Delta\theta <\theta_{\rm{e}}$)}

We approximate the emission as a top-hat jet with opening angle $\theta_{\rm j}=\thee$ and an initial Lorentz factor $\Gamma_0 \gg\Delta\theta^{-1}$. The afterglow  flux peaks at $t_{\rm peak}$ when an observer enters the beaming cone at $\Gamma \approx \Delta\theta^{-1}$, 
before seeing the entire emitting region and before it expands significantly. This case is similar to the previous one (initially on-axis), where $\Gamma_0$ is replaced with $\Delta\theta^{-1}$ (and the geometrical normalization factor is smaller by almost an order of magnitude):
\begin{eqnarray}
t_{\rm{peak}}\simeq 10 \,{\rm{d}}\,E_{\rm{k,iso,51}}^{1/3}n_{-4}^{-1/3}\Delta\theta_{-1}^{8/3}\ ,
   \label{peak time1}
\end{eqnarray}
and
\begin{eqnarray}
F_{\nu,\rm{peak}}(\rm{3~GHz})&\simeq& 250{\,\mu\rm{Jy}\,}E_{\rm{k,iso,51}}n_{-4}^{\frac{p+1}{4}}\epsilon_{\rm{B,-3}}^{\frac{p+1}{4}}\nonumber\\
&&\epsilon_{\rm{e},-1}^{p-1}\Delta\theta_{-1}^{2(1-p)}d_{\rm{40Mpc}}^{-2} \ .
   \label{maximum flux1}
\end{eqnarray}
These estimates imply that for $\Delta\theta=0.1$ rad and $E_{\rm{k,iso}} = 4 \times 10^{49}{\,\rm erg}$, the flux peaks at $\sim 10{\rm~\mu Jy}$ after about 3 days. At smaller angles ($\Delta\theta=0.1$ rad) the flux peaks  earlier and it is brighter, but after about 3 days it is similar to the one observed at $\Delta\theta=0.1\,\rm{rad}$. Since there are no observations before day 2, our limit implies that $E_{\rm{k,iso}}(\Delta\theta<0.1 {\,\rm rad}) \lesssim 4 \times 10^{49}{\,\rm erg}$. Above we found a similar limit for material that is initially on-axis with $\Gamma_0 > 5$. We therefore conclude that  the afterglow observations constraint  the isotropic equivalent kinetic energy carried by material with $\Gamma_0 > 5$ at an angle that is smaller than 0.1 rad away from the line-of-sight to be lower than about $3 \times 10^{49}{\,\rm erg}$.

\subsection{Far off-axis ($\Gamma_0^{-1}<\theta_{\rm{e}} < \Delta\theta $)}
Similarly to the near off-axis case we approximate the emission as a top-hat jet with an opening angle $\theta_{\rm j}=\thee$ and an initial Lorentz factor $\Gamma_0 \gg\Delta\theta^{-1}$. The difference is that here, by the time that the blast wave decelerates to $\Gamma\approx\Delta\theta^{-1}$ the Lorentz factor is smaller than $\thee^{-1}$ and we can see the entire emitting region. In addition at this point the emitting region has already expanded laterally significantly. This emission is similar to the one expected from a ``classical" off axis GRB jet \citep{Granot+2002,Nakar+2002}:
\begin{eqnarray}
t_{\rm{peak}}&\simeq& 90 \,{\rm{d}\,}E_{\rm{k,iso,51}}^{1/3}n_{-4}^{-1/3}\theta_{\rm{e,-1}}^{2/3}\Delta\theta_{-0.5}^{2} \ ,
\end{eqnarray}
and
\begin{eqnarray}
F_{\nu,\rm{peak}}&\simeq&4{\,\mu\rm{Jy}\,}E_{\rm{k,iso,51}}n_{-4}^{\frac{p+1}{4}}\epsilon_{\rm{B,-3}}^{\frac{p+1}{4}}\times \nonumber\\
&&\epsilon_{\rm{e},-1}^{p-1}\theta_{\rm{e,-1}}^2\Delta\theta_{-0.5}^{-2p}\nu_{\rm{3GHz}}^{\frac{1-p}{2}}d_{\rm{40Mpc}}^{-2}\ .
\end{eqnarray}
Note that the pre-factor here is different than the one used in  \cite{Granot+2002} and \cite{Nakar+2002}, since it is based on semi-analytic code that takes into account geometrical factors that were ignored in these papers.

\begin{figure}
\begin{center}
\includegraphics[width=60mm, angle=90]{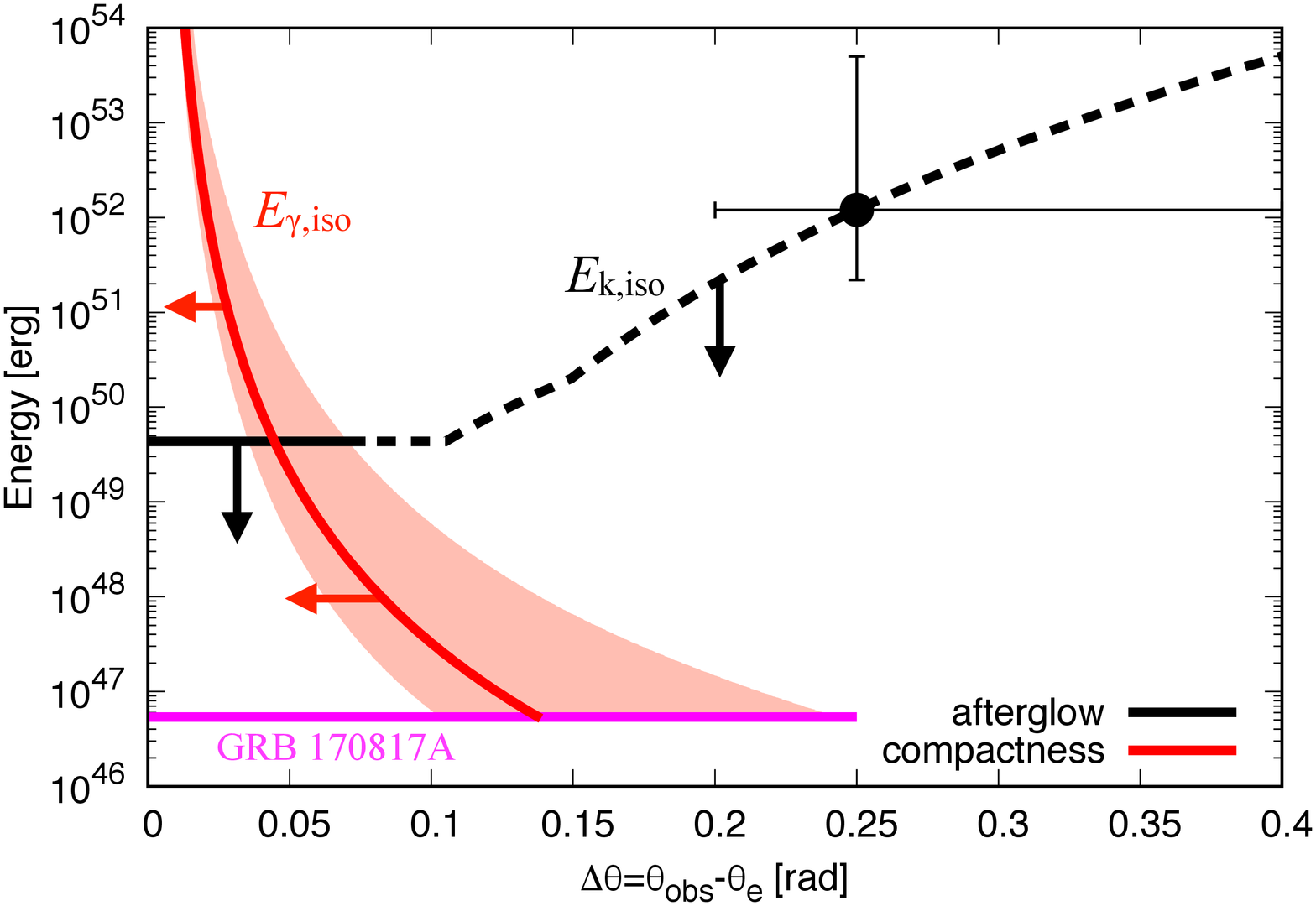}
\caption{Required conditions for the \gr emitting region.
The black curve shows the upper limit on the kinetic energy imposed by the afterglow.
The microphysics parameters are set as $n=10^{-4}{\,\rm{cm^{-3}}}$, $\epsilon_{\rm{e}}=10^{-1}$, and $\epsilon_{\rm{B}}=10^{-3}$ \citep{Mooley+2018b}.
The upper limits for the near off-axis (solid) and the far off-axis (dashed) cases are connected at $\Delta\theta=0.08\,\rm{rad}$.
The red curve shows the limit derived by the compactness argument for case ($i$) (see also Fig \ref{fig theta}).
The data point denotes the observed jet core's angular distance and the kinetic energy \citep{Mooley+2018b}.
The magenta line shows the observed $\gamma$-ray energy in GRB 170817A.
The region that emitted the $\gamma$-rays must lie in a narrow range above the observed magenta line, to the left of the compactness (red) curve and below the afterglow (black) line.
}
\label{fig energylimit}
\end{center}
\end{figure}

\section{Combined constraints on the source of GRB 170817A and comparison to sGRBs}\label{sec:GRBs} 
Fig. \ref{fig energylimit} combines the limits obtained in the previous two sections. The red line shows the compactness limit.  The $\gamma$-ray emitting region of GRB 170817A must lie to the left of this line. The black line marks the upper limit on $E_{\rm k,iso}(\Delta\theta)$. The isotropic equivalent afterglow kinetic energy must lie below this line. First, the combined constraints\footnote{ We use  the conservative assumption that $E_{\rm \gamma,iso} \lesssim E_{\rm k,iso}$, which holds unless the $\gamma$-ray efficiency is extreme even in GRB standards.} leave a very narrow allowed region for the source of GRB 170817A. 
The emitting region cannot be more than $5^\circ$ away from the line-of-site and the total isotropic equivalent energy carried by the emitting material cannot exceed $\sim3\times10^{49}$ erg, namely with a reasonable $\gamma$-ray efficiency ${\cal A} \lesssim 100$. 
The observed gamma-rays must have been emitted either on-axis (i.e., $\Delta\theta \lesssim\Gamma^{-1}$), or alternatively if there is a significant off-axis suppression then it is rather limited, corresponding to only  
 $\sim 10^{-3}$ of the jet core isotropic equivalent energy. Thus, if the core of the jet produced a regular sGRB towards an observer along its beaming cone, the sGRB  was brighter by several orders of magnitude than the emission from the region that produced the \gr that we observed, even if the luminosity that we saw was suppressed by off-axis effects.

\begin{figure}
\begin{center}
\includegraphics[width=60mm, angle=90]{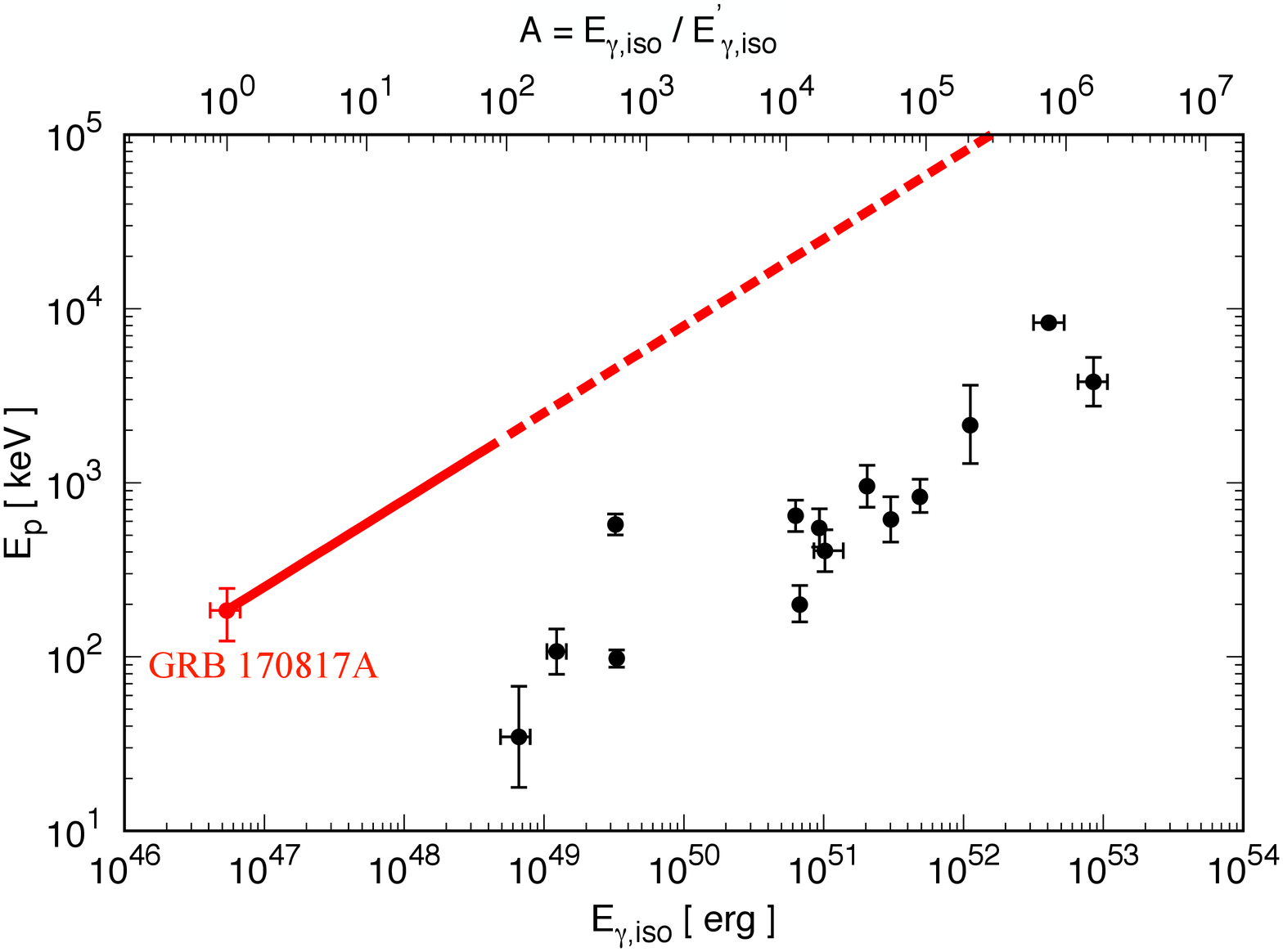}
\caption{The trajectory (for case $i$) on which  on-axis observables should be located  to produce the observed average properties of the prompt \gr in GRB 170817A. The dashed line marks regions with ${\cal A }> 100$ that are ruled out when compactness and afterglow considerations are taken into account. 
The black points show other observed sGRBs  from \citep{Wang+2017}.}
\label{fig amati}
\end{center}
\end{figure}

The observed \gr were $\sim 10^{-5}$ fainter than the putative sGRB emitted by the jet core. Our results show that even if off-axis effects were important in shaping GRB 170817A then an on-axis observer saw a signal that is at most $\sim 10^{-3}$ fainter than the sGRB produced by jet core.  Therefore there is no motivation to expect that off-axis effects played a significant role. Yet, it is interesting to ask how did  the on-axis emission looked like in that case. Taking the maximal value of ${\cal A} \sim 100$ results in $\sim 5 \times 10^{48}$ erg burst, which is weaker than  the faintest sGRB seen to date. The variability time scale of its pulses is $\sim 10$ ms and the average spectrum hardness is $\sim 2$ MeV. 
Fig. \ref{fig amati} shows the track that GRB 170817A can take on the $E_{\rm \gamma,iso}-E_{\rm p}$ plane as a result of off-axis effects. Clearly, there is no point along this track that resembles an sGRB we have seen before. Our conclusion is that the \gr that we observed were most likely emitted by a different emission mechanism than that of a regular GRB.

\section{Summary and Discussion}\label{sec:conclusions}

We derived two different constrains on the processes that took place in the neutron star merger event GW170817 and its EM counterparts. First we derived a constraint on the emission region that produced the prompt $\gamma$-rays. We have shown that  compactness implies a minimal  Lorentz factor ($\Gamma\sim 5$ for ``on-axis" source and larger for ``off-axis" ones) of the emitting region and a maximal angular distance from us to the source.  
In particular, this angular  separation should have been very small ($\lesssim 5^\circ$; see Fig. \ref{fig theta}) and it is much smaller than the estimated angular separation between our viewing angle and the core of the jet as found by \cite{Mooley+2018b} using the afterglow observations and by \cite{Mandel2018} and \cite{Finstad+2018} using the GW signal.

A second  independent constraint arises from the radio afterglow. 
The increase in the radio flux over the first 150 days implies  energy injection into the observed region during this period. This can be a result of either  a radial structure (more energetic shells move at lower velocity behind the shock front) or due to an angular structure (more energetic regions are located at a larger viewing angles and are observed only once they slow down) \citep{Nakar&Piran2018}. The observed superluminal motion of the centroid of the radio signal indicates that the dominant energy injection was angular. We set upper limits on the possible kinetic energy of the relativistic ($\Gamma \gtrsim 5$) matter as a function of the viewing angle.

The combination of the two constraints limits the possible conditions within the prompt $\gamma$-ray emitting region. The small angular distance between us and the $\gamma$-ray source, compared to the distance to the jet axis, suggests that the gamma-rays were emitted on-axis, and even if they were emitted off-axis then their on-axis luminosity was still very faint  and very hard ($E_{\gamma,\rm iso} \lesssim 5 \times 10^{48}$ erg; $E_{\rm p} \sim 2$ MeV) a combination that is unlike other sGRB observed so far.

\section*{acknowledgments}
This research is supported by the CHE-ISF I-Core center for excellence in Astrophysics. 
TM is supported by JSPS Overseas Challenge Program for Young Researchers and 
 by Grant-in-Aid for JSPS Research Fellow 17J09895. 
 TP is supported by an advanced ERC grant TReX and by the Templeton foundation.

\bibliographystyle{mnras}
\bibliography{reference_matsumoto}

\bsp
\label{lastpage}
\end{document}